# CURRENT OSCILLATIONS AND NEGATIVE RESISTANCES IN CROSSED CARBON NANOTUBES SUSPENDED OVER A DIELECTRIC TRENCH

M. Al Ahmad,[1] D. Dragoman,[2] M. Dragoman,[3] R. Plana,[1] J.-H. Ting,[4] F-Y. Huang,[5] and T.-L. Li [6]

[1]LAAS CNRS, 7 Avenue du Colonel Roche, 31077 Toulouse Cedex 4, France

[2]Univ. Bucharest, Physics Dept., P.O. Box MG-11, 077125 Bucharest, Romania

[3] National Research and Development Institute in Microtechnology, Str. Erou Iancu Nicolae 32B, 077190 Bucharest, Romania

[4]National Nano Device Laboratories (NDL) No.26, Prosperity Road I, Science-Based Industrial Park, Hsinchu, Taiwan 30078, R.O.C.

[5]National Central University, Dept. of Mechanical Engineering
300, Jung-da Road, Jung-li City, Taoyuan, Taiwan 32001, R.O.C.

[6]National Chia-Yi University, Dept. Applied Physics
300, Hsueh-Fu Road, Chia-Yi, Taiwan 60004, R.O.C.

ABSTRACT

An oscillatory dependence of the drain current on the drain voltage is found in a nanostructure consisting of two crossing semiconductor carbon nanotubes that are suspended over a dielectric trench, which is backed by a doped silicon substrate that acts as a gate. Alternating positive and negative differential resistance regions are generated as a function of the drain source values and can be slightly shifted by the gate voltage. Moreover, the negative differential resistance is retrieved in a large bandwidth, of up to 100 MHz, when the structure is excited with ac signals.

___________________________________________________________________________

a)  Corresponding author email: mircea.dragoman@imt.ro, mdragoman@yahoo.com

Crossed carbon nanotubes (CNTs) are very promising geometries for logic and memory applications at the nanoscale. For example, crossed nanotubes act as nanoelectro-mechanical system (NEMS) switches, the on- state corresponding to the case in which the two nanotubes come in contact due to a dc field applied on the gate [1]. Very large RAM memories could be fabricated based on this principle. Nanoscale electrical switches and logic applications were implemented with Y-shaped crossed nanotubes [2], which also display rectifying properties [3]. Useful reviews on crossed CNTs can be found in [4] and [5]. In practice, tens, hundreds or even thousands of similar crossed CNT structures are used for interconnection and/or logic applications, in which it is often desired to generate and/or amplify the signal. For example, a ring oscillator that generates a signal in the 10–50 MHz frequency range was realized by paralleling 10 FETs assembled on a single CNT [6].

We show in this paper that a single structure consisting of two crossed CNTs suspended over a trench between two electrodes is able to produce multiple negative differential resistance regions when a bias voltage is applied between the metallic contacts. Moreover, this negative differential resistance persists up to 100 MHz, the crossed CNTs being thus able to act as amplifiers or oscillators in nanoscale circuits. The negative differential resistance is the key element for oscillator devices consisting, in principle, of a negative differential resistance and a load.

The crossed CNT device is schematically displayed in Fig. 1 a-b, i.e. the top-view and the cross-view of the device. The fabrication of the structure and the SEM photos can be found in [7]. The distance between the S (source) and D (drain) electrodes is about 1.85 μm, the nanotubes with a diameter of 20 nm crossing at about 0.35 μm from the D contact at the crossing point is denoted by A in Fig. 1(a).

In this X-shaped configuration the two ends of the crossing CNTs are contacted by the same (source or drain) electrode, in contrast to a previous configuration where each termination of a crossed nanotube junction was contacted separately by an electrode [8]. Thus, in our case we have only two, S and D, electrodes instead of four distinct electrodes that appear in the configuration studied in [8]. Therefore, we expect the electrical characteristics of our device to be very different compared to those of crossed nanotube configurations reported before.

In Figs. 2(a) and 2(b), we have represented the drain-source current of the device for positive and negative drain-source voltages, respectively. From these figures it follows that we have drain current oscillations in the drain voltage range between −3 V and +3 V, which consist of alternating regions of positive and negative differential resistance with an almost constant peak-to-valley ratio of 2. Moreover, the drain current–drain voltage characteristic can be slightly shifted by a positive gate voltage only if the drain-source voltage is negative. These facts indicate an ambipolar conduction in an asymmetric device, the asymmetry originating from the quasiballistic character of the conduction and the asymmetry of the crossing point with respect to the S and D contacts.

In order to explain this unusual drain current–drain voltage characteristic, we have simulated the crossed CNT device considering each CNT as being formed from two coupled quantum wells. The first quantum well extends from the left contact up to the crossing point while the second is formed between the crossing point and the right side contact. In wide quantum wells, as is the situation in our case, there are numerous resonances/maxima in the quantum transmission, which generate oscillations in the drain current if the there is a sufficiently high contrast between the effective electron masses in the quantum well and the surrounding barriers [9], [10]. This is not our case, however, because there are no oscillations in the drain current–drain voltage characteristic in suspended non-crossed CNTs (these experiments are not shown here) fabricated in the same time, in the same conditions and located elsewhere on the wafer. Hence, the observed current oscillations in the crossed-CNT situation are due to the coupling between the two quantum wells. The simulation results presented in Fig. 3 correspond to electron effective masses in the Ti electrode, Ti/CNT barrier, quantum well, and crossed-CNT barrier of, respectively, 1. $m_0$, 1.1 $m_0$, 0.1 $m_0$ and 0.25 $m_0$, with $m_0$ the free electron mass. The widths of the Ti/CNT and crossed-CNT barriers were taken as 0.8 nm and 0.4 nm, respectively, their corresponding heights are 0.2 eV and 0.05 eV, and the potential profile was considered as step-like in the unbiased structures and linearly decreasing with the distance in the potential wells for the biased structure. We have used the Landauer formula for simulations. The simulations in Fig. 3 show the same number of current maxima as the experimental curve, which confirm that the mechanism responsible for current oscillations is the coupling between the two quantum wells. The low current value (of pA) results from the fact that most of the current generated by the external source flows through the low-resistivity *p*-Si gate layer.

The agreement between the simulations and the experimental results is quite unexpected since the length of the device, and in particular the length of the second quantum well is above the mean free path length of CNTs at room temperature. This agreement implies that the transport is quasiballistic at room temperature, so that the transmission characteristic of the wide second well is still modulated by the first, much narrower, quantum well. This quasiballistic transport characteristic also explains the odd gate voltage effect on the drain current–drain voltage curve: when the electrons flow from the left contact to the one at the right, the gate voltage effect on the current-voltage characteristics is that expected from ballistic transport considerations, since the electron propagation through the first quantum well encountered is ballistic and this ballistic transmission is that which modulates the transmission through the wider quantum well. On the contrary, when the electrons flow from the right contact, they arrive at the second contact as quasiballistic electrons, and hence less prone to be affected by the gate voltage; the transmission peaks are much less shifted by the gate voltage because narrower quantum well characteristic modulates the transmission of quasiballistic instead of ballistic electrons.

We have further studied the behaviour of the negative differential resistance regions under ac signals excitation. We found that the negative differential resistance is still present up to 100 MHz when the structure is excited with an ac generator between the gate and the ground, and the output is collected from the drain contact. This result is presented in Fig. 4(a) for an ac voltage amplitude of 0.5 V and in the frequency range of 100 kHz–100 MHz. In Fig. 4(b) we have represented the resistance of the structure at 1 MHz and at various amplitude excitation levels. We can see that the differential resistance is always negative from 0.35 V, up to 1 V amplitude. The reactance of the structure displays a capacitive behaviour in the amplitude and frequency ranges of the experiments.

In conclusion, a crossed CNT structure is showing unexpected physical phenomena, such as oscillations in the drain current as a function of the drain voltage, and negative differential resistances up to 100 MHz due to the coupling between quantum wells formed by the crossed CNTs. These effects can be used to generate or even amplify currents in nanoelectronic platforms containing many nanotubes, in nanoradios or multi-valued logics.

FIGURE CAPTIONS

Fig. 1   The crossed CNT device. (a) Top -view, and (b) cross-section.

Fig. 2   Drain current versus (a) negative and (b) positive drain voltage values, in the presence and absence of a gate voltage.

Fig. 3   Simulated drain current versus drain voltage in the absence of a gate voltage.

Fig. 4   The negative resistance as (a) a function of frequency, and (b) a function of the amplitude level.

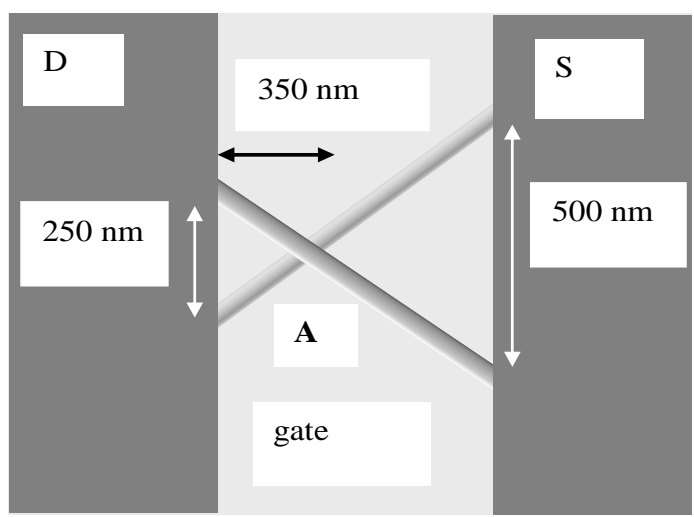

**(a)**

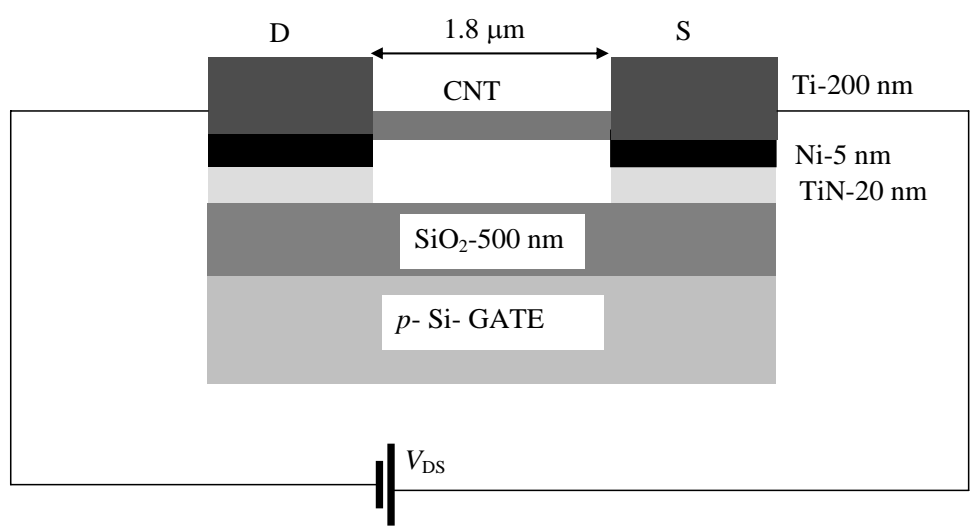

**(b)**

Fig. 1

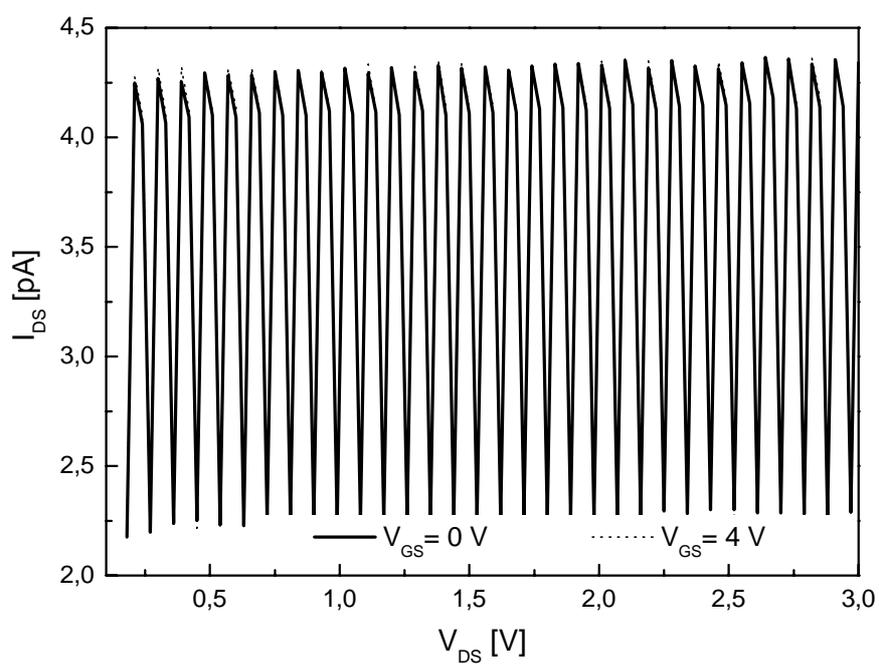

**(a)**

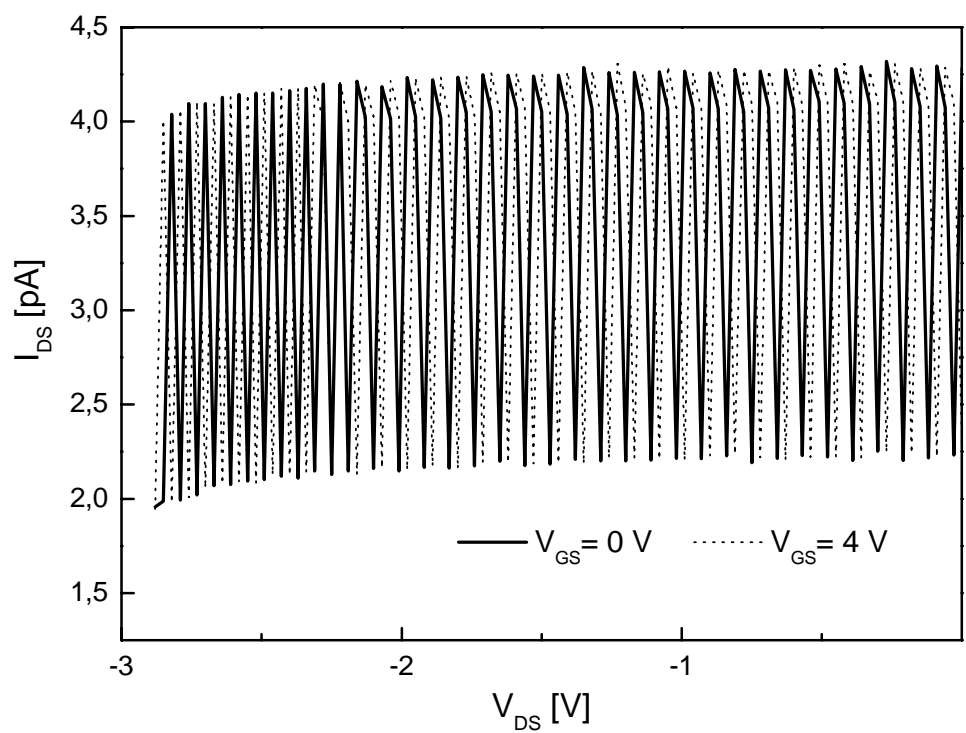

**(b)**

Fig. 2

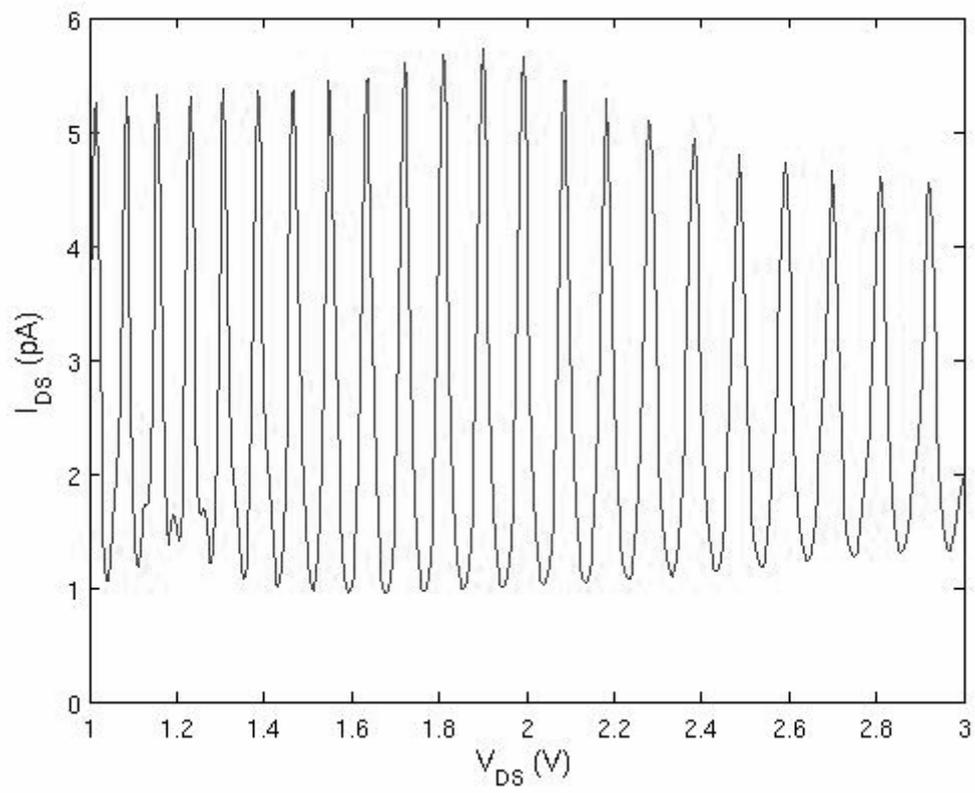

Fig. 3

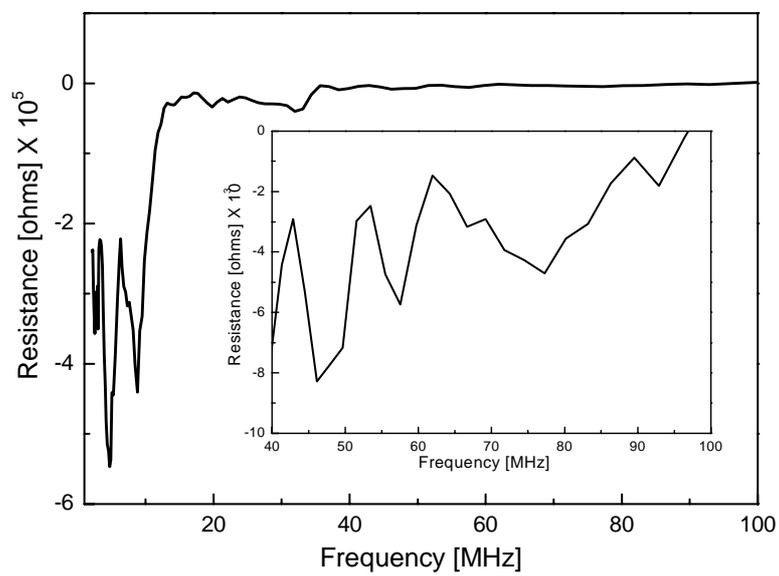

**(a)**

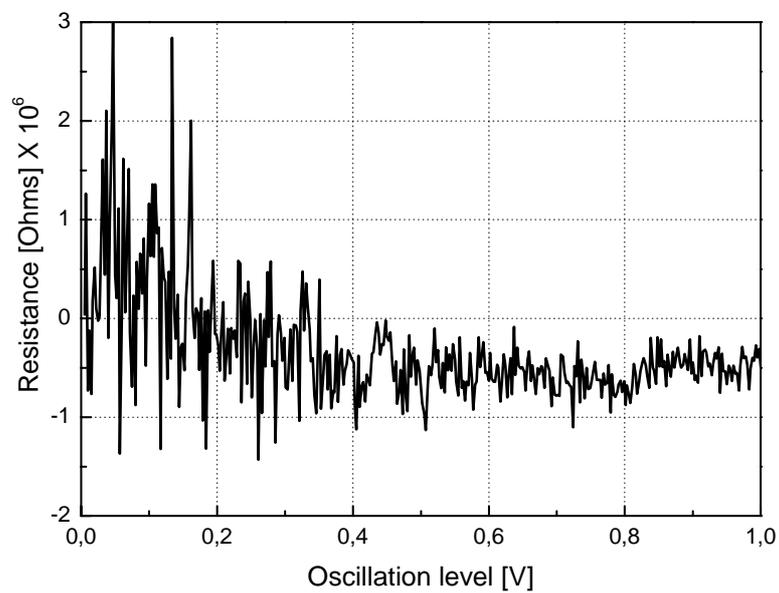

**(b)**

Fig. 4